\documentclass[twocolumn,twoside]{IEEEtran}
\usepackage{mathpazo}
\usepackage{times}
\usepackage{color}
\usepackage{amsmath}
\usepackage{amsfonts}
\usepackage{latexsym}
\usepackage{amssymb}
\usepackage{cite}

\usepackage{upref}
\usepackage{theorem}
\usepackage{graphicx}
\usepackage{psfrag}

\theoremstyle{plain}
\theorembodyfont{\normalfont\slshape}

\newtheorem{thm}{Theorem$\!$}
\newenvironment{theorem}
{\begin{thm}\hspace*{-1ex}{\bf.}}{\end{thm}}

\newtheorem{lem}[thm]{Lemma$\!$}
\newenvironment{lemma}{\begin{lem}\hspace*{-1ex}{\bf.}}{\end{lem}}

\newtheorem{prop}[thm]{Proposition$\!$}

\newtheorem{cor}[thm]{Corollary$\!$}
\newenvironment{corollary}{\begin{cor}\hspace*{-1ex}{\bf.}}{\end{cor}}

\newtheorem{defn}[thm]{Definition$\!$}
\newenvironment{definition}{\begin{defn}\hspace*{-1ex}{\bf.}}{\end{defn}}

\newtheorem{xmpl}[thm]{Example$\!$}
\newenvironment{example}{\begin{xmpl}\hspace*{-1ex}{\bf.}}{\hfill$\Box$\end{xmpl}}

\newtheorem{cnstr}{Construction$\!$}

\newenvironment{construction}{\begin{cnstr}\hspace*{-1ex}{\bf.}}{\hfill$\Box$\end{cnstr}}

\newcounter{enumrom}
\renewcommand{\theenumrom}{(\roman{enumrom})}

\makeatletter
\renewcommand{\@endtheorem}{\endtrivlist}
\makeatother

\makeatletter
\renewcommand{\thefigure}{{\@arabic\c@figure}}
\renewcommand{\fnum@figure}{{\bf Figure\,\thefigure}}
\makeatother

\newcommand{\cG}{\mathcal{G}}

\newcommand{\mathset}[1]{\left\{#1\right\}}
\newcommand{\abs}[1]{\left|#1\right|}
\newcommand{\ceilenv}[1]{\left\lceil #1 \right\rceil}
\newcommand{\floorenv}[1]{\left\lfloor #1 \right\rfloor}
\newcommand{\parenv}[1]{\left( #1 \right)}

\newcommand{\be}[1]{\begin{equation}\label{#1}}
\newcommand{\ee}{\end{equation}}

\renewcommand{\leq}{\leqslant}
\renewcommand{\ge}{\geqslant}
\renewcommand{\geq}{\geqslant}

\renewcommand{\Bbb}{\mathbb}

\newcommand{\Cref}[1]{Co\-ro\-lla\-ry\,\ref{#1}}

\renewcommand{\Bbb}{\mathbb}

\newcommand{\gf}{{\mathrm{GF}}}

\newcommand{\N}{{\Bbb N}}
\newcommand{\R}{{\Bbb R}}
\newcommand{\Z}{{\Bbb Z}}

\newcommand{\dmin}{d_{\mathrm{min}}}
\DeclareMathOperator{\id}{Id}
\DeclareMathOperator{\rank}{rank}
\newcommand{\ccap}{\mathsf{cap}}
\newcommand{\ccaps}{\mathsf{cap}_{\mathrm{sys}}}

\outer\def\proclaim #1. #2\par{\medbreak
 \noindent{\bf#1.\enspace}{\sl#2\par}%
 \ifdim\lastskip<\medskipamount \removelastskip\penalty55\medskip\fi}

\begin{document}


\title{\Huge\bf Systematic Error-Correcting Codes \\for Rank Modulation}

\author{\large
Hongchao~Zhou,~\IEEEmembership{Member,~IEEE},
Moshe~Schwartz,~\IEEEmembership{Senior Member,~IEEE},\\
Anxiao~(Andrew)~Jiang,~\IEEEmembership{Senior Member,~IEEE},
and Jehoshua~Bruck~\IEEEmembership{Fellow,~IEEE}%
\thanks{This work was supported in part by a United States - Israel
  Binational Science Foundation (BSF) grant 2010075, a National Science
Foundation (NSF) CAREER Award CCF-0747415, an NSF grant CCF-1217944,
and an NSF grant CCF-1218005.}%
\thanks{The material in this paper was presented in part at the 2012 IEEE International Symposium on Information Theory.}%
\thanks{Hongchao Zhou is with the Research Laboratory of Electronics, Massachusetts Institute of Technology, Cambridge, MA 02139, USA (e-mail: hongchao@mit.edu).}%
\thanks{Moshe Schwartz is with the Department
   of Electrical and Computer Engineering, Ben-Gurion University of the Negev,
   Beer Sheva 8410501, Israel
   (e-mail: schwartz@ee.bgu.ac.il).}%
\thanks{Anxiao (Andrew) Jiang is with the Department of Computer Science and Engineering,
  Texas A\&M University, College Station, TX 77843-3112, U.S.A.
  (e-mail: ajiang@cse.tamu.edu).}%
\thanks{Jehoshua Bruck is with the Department
  of Electrical Engineering, California Institute of Technology, Pasadena, CA 91125, USA (e-mail: bruck@paradise.caltech.edu).}%
}

\maketitle

\begin{abstract}
The rank-modulation scheme has been recently proposed for efficiently
storing data in nonvolatile memories. Error-correcting codes are
essential for rank modulation, however, existing results have been
limited. In this work we explore a new approach, \emph{systematic
  error-correcting codes for rank modulation}. Systematic codes have
the benefits of enabling efficient information retrieval and
potentially supporting more efficient encoding and decoding
procedures. We study systematic codes for rank modulation under
Kendall's $\tau$-metric as well as under the $\ell_\infty$-metric.

In Kendall's $\tau$-metric we present $[k+2,k,3]$-systematic codes for
correcting one error, which have optimal rates, unless systematic
perfect codes exist. We also study the design of
multi-error-correcting codes, and provide two explicit constructions,
one resulting in $[n+1,k+1,2t+2]$ systematic codes with redundancy at
most $2t+1$. We use non-constructive arguments to show the existence
of $[n,k,n-k]$-systematic codes for general parameters. Furthermore,
we prove that for rank modulation, systematic codes achieve the same
capacity as general error-correcting codes.

Finally, in the $\ell_\infty$-metric we construct two $[n,k,d]$
systematic multi-error-correcting codes, the first for the case of
$d=O(1)$, and the second for $d=\Theta(n)$. In the latter case, the
codes have the same asymptotic rate as the best codes currently known
in this metric.
\end{abstract}

\begin{IEEEkeywords}
flash memory, rank modulation, error-correcting codes, permutations,
metric embeddings, Kendall's $\tau$-metric, $\ell_\infty$-metric, systematic
codes
\end{IEEEkeywords}


\section{Introduction}

\IEEEPARstart{T}{he} rank-modulation scheme has been recently proposed
for storing data efficiently and robustly in nonvolatile memories
(NVMs) \cite{JiaMatSchBru09}. Its applications include flash memories
\cite{CapGolOliZan99}, which are currently the most widely used family
of NVMs, and several emerging NVM technologies, such as phase-change
memories \cite{Bur10}. The rank-modulation scheme uses the relative
order of cell levels to represent data, where a cell level denotes a
floating-gate cell's threshold voltage for flash memories and denotes
a cell's electrical resistance for resistive memories (such as
phase-change memories). Consider $n$ memory cells, where for
$i\in[n]=\mathset{1,2,\dots,n}$, let $c_{i}\in \R$ denote the level of
the $i$th cell. It is assumed that no two cells have the exact same
level, which is easy to realize in practice. Let $S_{n}$ denote the
set of all $n!$ permutations over $[n]$. The $n$ cell levels induce a
permutation $[f_{1},f_{2},\dots,f_{n}]\in S_{n}$, where
$c_{f_{1}}>c_{f_{2}}>\dots > c_{f_{n}}$. The rank-modulation scheme
uses such permutations to represent data. It enables memory cells to
be programmed efficiently and robustly, from lower levels to higher
levels, without the risk of over-programming. It also makes it easier
to adjust cell levels when noise appears without erasing cells, and
makes the stored data more robust to asymmetric errors that change
cell levels in the same direction
\cite{JiaMatSchBru09,JiaSchBru10,TamSch10}.

Error-correcting codes are essential for data
reliability. Intuitively, an error-correcting code is a set of
elements in a metric space, no two of which are too close together
under its distance measure. In the case of rank modulation, the space
is $S_n$. As for the distance measure, it is usually chosen in such
a way that small (common) errors in the physical medium correspond to
a small distance in the metric space. In the context of rank
modulation for NVMs, the two most studied distance functions are
Kendall's $\tau$-distance, and the $\ell_\infty$-distance. It was
suggested in \cite{JiaSchBru10} that small charge-constrained errors
correspond to a small distance in Kendall's $\tau$-metric. In
contrast, in \cite{TamSch10} it was shown that small limited-magnitude
errors correspond to a small $\ell_\infty$-distance.

Some results are known on error-correcting codes for rank modulation
equipped with Kendall's $\tau$-distance. In \cite{JiaSchBru10}, a
one-error-correcting code is constructed based on metric embedding,
whose size is provably within half of the optimal size. In
\cite{BarMaz10}, the capacity of rank modulation codes is derived for
the full range of minimum distance between codewords, and the
existence of codes whose sizes are within a constant factor of the
sphere-packing bound for any fixed number of errors is shown. Some
explicit constructions of error-correcting codes have been proposed
and analyzed in \cite{MazBarZem11}. We also mention that the Ulam
metric has been suggested as a generalization of Kendall's
$\tau$-metric and was recently studied in the context of
error-correcting codes in \cite{FarSkaMil13}.

There has also been some work on error-correcting codes for rank
modulation equipped with the $\ell_{\infty}$-distance. In
\cite{TamSch10,KloLinTsaTze10} some general constructions and bounds
were given. A relabeling scheme, improving the distance of codes was
suggested in \cite{TamSch12}. Several counting problems, mainly
concerning ball size under the $\ell_\infty$-metric, and optimal
anticodes, were studied in \cite{Sch09,Klo09,Klo11,SchTam11}.

In this paper, we study \emph{systematic} error-correcting codes for
rank modulation as a new approach for code design. In the more common
error-correcting setting over vectors equipped with the Hamming
distance function, an $[n,k]$-systematic code is a subset of
length-$n$ vectors whose projection onto a given set of $k$
coordinates has all possible length-$k$ vectors appearing exactly
once. These $k$ positions are referred to as the \emph{information
  symbols}, whereas the rest of the positions are called
\emph{redundancy symbols}. If the code is linear, it is well known
(for example, see \cite{MacSlo78}) that any code has an equivalent
code with the same parameters that is also systematic. We shall be
interested in the analog of systematic codes in the space of
permutation with either Kendall's $\tau$-distance or the
$\ell_\infty$-distance. Compared with the existing constructions of
error-correcting codes for rank modulation, systematic codes have the
benefit that they support efficient data retrieval, because when there
is no error (or when error correction is not required), data may be
retrieved by only reading the information symbols. Since every
permutation induced by the information symbols appears in exactly one
codeword, the codewords can be mapped efficiently to data (and vice
versa) via enumerative source coding (e.g., by ordering permutations
lexicographically) \cite{Cov73,MarStr07}. In addition, the encoding
algorithm of the error-correcting code can potentially be made very
efficient by defining the positions of the redundant cells in the
permutation as a function of the corresponding positions of the
information cells.

We study the design of systematic codes, and analyze their
performance. In Kendall's $\tau$-metric we present families of
$[k+2,k]$-systematic codes for correcting a single error. We show that
they have optimal parameters among systematic codes, unless
\emph{perfect} systematic one-error-correcting codes, which meet the
sphere-packing bound, exist. We also study the design of systematic
codes that correct multiple errors, prove the existence of codes with
minimum distance $n-k$ for any $2\leq k < n$, as well as give a
construction for a wide range of parameters. Furthermore, we prove
that systematic codes have the same capacity as general
error-correcting codes. This result establishes that, asymptotically,
systematic codes are as strong in their error-correction capability as
general codes. We also consider the $\ell_\infty$-metric, and provide
a general construction for systematic code whose asymptotic rate
equals that of the best codes currently known.

The rest of the paper is organized as follows. In Section
\ref{sec:terms} we provide the basic notation and definitions used
throughout the paper. In Section \ref{sec:kt} we study systematic
codes in Kendall's $\tau$-metric. We turn in Section \ref{sec:linf} to
explore systematic codes in the $\ell_\infty$-metric. We conclude in
Section \ref{sec:conclusion} and present some open problems.


\section{Notation and Definitions}
\label{sec:terms}

Let $[n]=\mathset{1,2,\dots,n}$, and $S_n$ denote the set of
permutations over $[n]$. A permutation $f\in S_n$ is represented in
single-line notation by $f=[f_1,f_2,\dots,f_n]$, where $f(i)=f_i$. We
also denote the identity permutation by $\id=[1,2,\dots,n]$. Finally,
we denote by $f^{-1}$ the inverse of the permutation $f$, i.e., the
permutation sending $f(i)$ to $i$.

Consider a metric over the permutations $S_n$ with a distance function
$d:S_n\times S_n\to\N\cup\mathset{0}$. An $(n,M,\dmin)$-code is a
subset $C\subseteq S_n$ such that $\abs{C}=M$, and $d(f,g)\geq \dmin$
for all $f,g\in C$, $f\neq g$. We say $M$ is the \emph{size} of the
code, and $\dmin$ is the minimum distance of the code.

In this work we shall consider two distance functions: Kendall's
$\tau$-distance, and the $\ell_\infty$-distance. The latter
$\ell_\infty$-distance function is easily defined for all $f,g\in S_n$
by
\[d_\infty(f,g)=\max\mathset{\abs{f(i)-g(i)} ~|~ i\in [n]}.\]
For the former, Kendall's $\tau$-distance function, assume $f\in S_n$
is some permutation. An \emph{adjacent transposition} on $f$ switches
the values of $f(i)$ and $f(i+1)$ for some $i\in [n-1]$.
Kendall's $\tau$-distance \cite{KenGib90} between $f$ and $g$, denoted
$d_K(f,g)$, is defined as the minimal number of adjacent
transpositions required to transform $f$ into $g$. This is sometimes
also called the \emph{bubble-sort distance}.

We recall that in the rank-modulation scheme we have $n$ memory cells
labeled by $[n]$, and the level of the $i$th cell is denoted by
$c_i\in\R$. Assume $c_{i_1}>c_{i_2}>\dots>c_{i_n}$, then the
permutation stored by the $n$ cells is $[i_1,i_2,\dots,i_n]\in S_n$
(see \cite{JiaMatSchBru09}). Assume a permutation $f\in S_n$ was
stored, but a distorted version of it, $g\in S_n$, was eventually
read. It was noted in \cite{JiaSchBru10} that small charge-constrained
errors translate to small Kendall's $\tau$-distance, denoted
$d_K(f,g)$. In contrast, it was suggested in \cite{TamSch10,TamSch12},
that limited-magnitude errors translate to small
$\ell_\infty$-distance on the inverse permutation, denoted
$d_\infty(f^{-1},g^{-1})$. This difference between storing the
permutation or its inverse will play a role in defining two versions
of systematic codes.

In order to define systematic codes we need to define two types of
projections. Let $A=\mathset{a_1,a_2,\dots,a_m}\subseteq [n]$ be any
subset, $a_1<a_2<\dots<a_m$. For any permutation $f\in S_n$, we define
$f|_A$ to be the permutation in $S_m$ that preserves the relative
order of the sequence $f(a_1),f(a_2), \dots,f(a_m)$. Intuitively, to
compute $f|_A$ we keep only the \emph{coordinates} of $f$ that appear
in $A$, and then relabel the entries to $[m]$ while the keeping
relative order. In a similar fashion we define
\[f|^A= \parenv{f^{-1}|_A}^{-1}.\]
Intuitively, to calculate $f|^A$ we keep only the \emph{values}
of $f$ from $A$, and then relabel the entries to $[m]$ while
keeping relative order.

\begin{example}
Let $n=6$ and consider the permutation
\[f=[6,1,3,5,2,4]\in S_6.\]
We take $A=\mathset{3,5,6}$. We then have
\[f|_A=[2,1,3],\]
since we keep positions $3$, $5$, and $6$, of $f$, giving us
$[3,2,4]$, and then relabel these to get $[2,1,3]$.

Similarly, we have
\[f|^A=[3,1,2],\]
since we keep the values $3$, $5$, and $6$, of $f$, giving us
$[6,3,5]$, and then relabel these to get $[3,1,2]$.
\end{example}

We are now in a position to define systematic codes in two different
ways, depending on the metric.

\begin{definition}
An $[n,k,d]$ systematic code, $C$, for Kendall's $\tau$-metric, is an
$(n,k!,d)$ code such that
\[\mathset{ \left. f|^{[k]} ~\right|~ f\in C} = S_k.\]
We call $[k]$ the \emph{information symbols} of the code, and
$\mathset{k+1,k+2,\dots,n}$ the \emph{redundancy symbols} of the code.
\end{definition}

Intuitively, if we have an $[n,k,d]$-systematic code in Kendall's
$\tau$-metric, reading just the levels of the first $k$ cells and
comparing them, enables us to ascertain the relative positions of the
values $1,2,\dots,k$ in the stored permutation, and there is a unique
codeword with this relative ordering. More precisely, assume a
codeword $f\in C$ is stored. If the levels we read from the first $k$
cells are $c_1,c_2,\dots,c_k$, and their ordering is
$c_{i_1}>c_{i_2}>\dots>c_{i_k}$, then $i_1$ appears before $i_2$ in
the codeword, appearing before $i_3$, and so on, until $i_k$, i.e.,
$f^{-1}(i_1)<f^{-1}(i_2)<\dots<f^{-1}(i_k)$.

In contrast, in the setting of limited-magnitude errors and the
$\ell_\infty$-metric \cite{TamSch10}, the \emph{inverse} of the
permutation read from the cells is protected by an error-correcting
code. Thus, if $g$ is the codeword we want to store, we would
physically write its inverse $g^{-1}$ to the cells using the
rank-modulation scheme. Then, reading just the levels of the first $k$
cells, $c_1,c_2,\dots,c_k$, gives us the relative ordering of
$g(1),g(2),\dots,g(k)$. This motivates the following definition.

\begin{definition}
An $[n,k,d]$ systematic code, $C$, for the $\ell_\infty$-metric, is an
$(n,k!,d)$ code such that
\[\mathset{ \left. g|_{[k]} ~\right|~ g\in C} = S_k.\]
We call $[k]$ the \emph{information coordinates} of the code, and
$\mathset{k+1,k+2,\dots,n}$ the \emph{redundancy coordinates} of the
code.
\end{definition}


\section{Systematic Codes in Kendall's $\tau$-Metric}
\label{sec:kt}

This section is devoted to the study of systematic codes in Kendall's
$\tau$-metric. In Section \ref{sec:ktprem} we introduce further
notation and some useful lemmas. In Section \ref{sec:ktone} we study
systematic one-error-correcting codes. We turn, in Section
\ref{sec:ktmulti}, to the case of general systematic error-correcting
codes. Finally, in Section \ref{sec:ktcap}, we analyze the capacity of
systematic codes.

\subsection{Preliminaries}
\label{sec:ktprem}

We let $\Z_n$ denote the set of integers $\mathset{0,1,\dots,n-1}$, as
well as the additive group over these integers with addition modulo
$n$. It is well known (see \cite{JiaSchBru10}, and references therein)
that there is a one-to-one correspondence between the permutations of
$S_n$ and \emph{factoradic} representations, which are mixed-radix
vectors from
\[\Z_n!=\Z_1\times \Z_2 \times \dots \times \Z_{n-1} \times \Z_n.\]
Let $f\in S_n$ be any permutation. The factoradic representation
corresponding to $f$ is a vector $v=[v_1,\dots,v_n]\in Z_n!$ such
that $v_i\in \Z_{i}$ equals
\[v_i=\abs{\mathset{ j ~\left|~ \text{$j < i$ and $f^{-1}(j)>f^{-1}(i)$}\right.}},\]
i.e., $v_i$ counts the number of elements of lesser value than $i$,
but which appear to the right of $i$ in the vector
$f=[f_1,f_2,\dots,f_n]$. We note that $v_1=0$ always, and is thus
redundant in the representation, but we keep it to make the notation
simpler.  From now on, we denote the factoradic representation of
$f\in S_n$ by $\Phi(f)\in\Z_n!$, and the $i$th element of $\Phi(f)$ by
$\Phi(f)_i$.

We now crucially observe that, in a systematic scheme, setting the
levels of the first $k$ cells determines exactly the first $k$ entries
of the factoradic representation of the permutation stored by the $n$
cells. This is true regardless of the levels of the last $n-k$ cells.
More succinctly, for any $f\in S_n$, and for all $1\leq i\leq k\leq n$,
\[\Phi(f|^{[k]})_i = \Phi(f)_i.\]

\begin{example}
Let $n=6$ and $k=4$. Take
\[f=[6,1,3,2,5,4]\in S_n.\]
We then have
\[\Phi(f)=[0,0,1,0,1,5],\]
as well as
\[f|^{[k]}=[1,3,2,4] \qquad \text{and} \qquad \Phi(f|^{[k]})=[0,0,1,0].\]
We observe that the first $k$ coordinates of $\Phi(f)$ and $\Phi(f|^{[k]})$
are the same.
\end{example}

Another well-known fact (used by \cite{JiaSchBru10,BarMaz10}) is the
following metric embedding:
\begin{equation}
\label{eq:distl1}
d_K(f,g) \geq d_1(\Phi(f),\Phi(g)) = \sum_{i=1}^n \abs{\Phi(f)_i-\Phi(g)_i},
\end{equation}
where $d_K$ is Kendall's $\tau$-distance, and $d_1$ is the
$\ell_1$-distance. The following lemma gives a more refined version of
\eqref{eq:distl1}, taking into account the partition into information
symbols and redundancy symbols.

\begin{lemma} \label{theorem1}
Given $f,g\in S_n$, and $1\leq k\leq n$,
\[d_K(f,g)\geq d_K(f|^{[k]},g|^{[k]})+\sum_{i=k+1}^{n}\abs{\Phi(f)_i-\Phi(g)_i}.\]
\end{lemma}
\begin{IEEEproof}
The proof is by induction on $r=n-k$. As the base case, the inequality
is clearly satisfied for $r=0$, i.e., $n=k$. Now consider the
inductive step. Suppose that the inequality holds for some
$r-1=n-k-1$, and we will now show that it also holds for $r=n-k$.
	
Consider a sequence of $d_K(f,g)$ adjacent transpositions that changes
the permutation $f$ into the permutation $g$. Of these transpositions,
assume that $\alpha$ adjacent transpositions involve the integer $n$,
and $\beta$ adjacent transpositions do not involve $n$. Clearly,
\[d_K(f,g)=\alpha+\beta.\]

Since the integer $n$ needs to be moved from position
$n-\Phi(f)_n$ to position $n-\Phi(g)_n$, we get
\[\alpha \geq \abs{\Phi(f)_n-\Phi(g)_n}.\]
Note that those adjacent transpositions that involve $n$ do not change
the relative order of the integers $[n-1]$ in the permutation. Thus,
to transform the integers $[n-1]$ from their relative order in $f$ to
their relative order in $g$, by the induction assumption, we
get 
\[\beta\geq
d_K(f|^{[k]},g|^{[k]})+\sum_{i=k+1}^{n-1}\abs{\Phi(f)_i-\Phi(g)_i}.\]
That leads to the conclusion.
\end{IEEEproof}

\begin{example}\label{example:3}
Let $n=3$ and $k=2$ and consider
\[ f=[1,3,2] \qquad \text{and} \qquad g=[2,1,3].\]
In this case, the inequality of Lemma \ref{theorem1} becomes an
equality since
\begin{multline*}
d_K([1,3,2],[2,1,3]) = 2 \\
= 1 + \abs{1-0} = d_K([1,2],[2,1])+\abs{\Phi(f)_3-\Phi(g)_3}.
\end{multline*}
The equality, however, does not always hold. For instance, if
\[ f'=[1,3,2] \qquad \text{and} \qquad g'=[2,3,1],\]
we get
\begin{multline*}
d_K([1,3,2],[2,3,1]) = 3 \\
> 1 + \abs{1-1} = d_K([1,2],[2,1])+\abs{\Phi(f)_3-\Phi(g)_3}.
\end{multline*}
\end{example}

We now present an inequality for ball sizes in $S_{n}$, which will be
useful for the analysis of systematic codes. Given a permutation
$f\in S_{n}$, the ball of radius $r$ centered at
$f$, is defined by
\[\mathfrak{B}_r(f)=\mathset{g\in S_{n} ~|~ d_K(f,g)\leq r},\]
for any $0\leq r\leq \binom{n}{2}$. We recall that the maximum
distance for any two permutations in $S_{n}$ is $\binom{n}{2}$ (for
example, see \cite{JiaSchBru10}). A simple relabeling argument
suffices to show that the size of a ball does not depend on the choice
of its center. Therefore, we will use $\abs{\mathfrak{B}_r(n)}$ to denote
$\abs{\mathfrak{B}_r(f)}$ for any $f\in S_{n}$.

An exact expression for $\abs{\mathfrak{B}_r(n)}$ is known
\cite{Knu98}. However, for our purposes, we will use the inequality of
the following lemma.

\begin{lemma}
\label{lemma2}
For all $n\geq 1$ and $0\leq r\leq \binom{n}{2}$,
\[\abs{\mathfrak{B}_r(n)}\leq \binom{n+r-1}{n-1}.\]
\end{lemma}
\begin{IEEEproof}
Since the center of a ball does not affect its size, consider the ball
centered at the identity, $\mathfrak{B}_r(\id)$. It follows from
\eqref{eq:distl1} that
\begin{equation}
\label{eq:ballbound}
\abs{\mathfrak{B}_r(\id)} \leq \abs{\mathset{f \in S_n ~|~
    d_1(\Phi(f),\Phi(\id))\leq r}}.
\end{equation}
Since, conveniently, $\Phi(\id)$ is the all-zero vector, we have
for any $f\in S_n$ that
\[d_1(\Phi(f),\Phi(\id))=\sum_{i=1}^n \Phi(f)_i.\]
We further note that $\Phi(f)_1=0$ always.

Thus, the right-hand side of \eqref{eq:ballbound} is upper bounded by
the number of non-negative-integer vectors of length $n-1$ whose entry
sum is at most $r$.  This is easily seen to be the same as the number
of ways $r$ identical balls can be thrown into $n$ non-identical bins,
and hence,
\[\abs{\mathfrak{B}_r(n)}\leq \binom{n+r-1}{n-1}.\]
\end{IEEEproof}

\subsection{Systematic One-Error-Correcting Codes}
\label{sec:ktone}

We start by presenting two constructions for systematic $[k+2,k,3]$
codes, capable of correcting a single error. The first construction
uses a direct manipulation of the permutations to construct the
codewords, and is somewhat restricted in its choice of parameters. In
contrast, the second construction uses a metric embedding technique,
and applies to all parameters. We then show the codes are optimal
unless perfect one-error-correcting codes exist.

\begin{construction}
\label{constr:1}
Let $k\ge 3$ be an integer such that either $k$ or $k+1$ is a prime.
For any $f\in S_k$, and $j\geq 1$ an integer, we define the following
function:
\begin{equation}
\label{eq:rho}
\rho_j(f) = \parenv{\sum_{i=1}^k (2i-1)^j f(i)} \bmod  m,
\end{equation}
where $m=k$ if $k$ a prime and $m=k+1$ if $k+1$ is a prime.

We construct the code
\[C = \mathset{f\in S_{k+2} ~\left|~ \text{$\Phi(f)_{k+j}=\rho_j(f|^{[k]})$ for
all $j\in[2]$}\right.}.\]
\end{construction}

\begin{theorem}
\label{thm:oneCode}
The code $C$ from Construction \ref{constr:1} is a systematic
$[k+2,k,3]$ code in Kendall's $\tau$-metric.
\end{theorem}
\begin{IEEEproof}
We easily observe that the information symbols $[k]$ are unconstrained, and so
\[\mathset{\left. f|^{[k]} ~\right|~ f\in C} = S_k.\]
Furthermore, since a choice of the order of the information symbols
determines the position of the two redundancy symbols uniquely, we
also have $\abs{C}=k!$.

It now only remains to show that the minimum distance of $C$ is
$3$. We know that either $k$ is a prime, or $k+1$ is a prime. Let us
first handle the former case. Let $f,g\in C$ be two codewords, $f\neq
g$. We divide our proof into three cases, depending on
$d_K(f|^{[k]},g|^{[k]})$.

\paragraph{Case 1} $d_K(f|^{[k]},g|^{[k]})=1$. In this case, we can write 
\begin{align*}
f|^{[k]}&=[a_1,a_2,\dots,a_i,a_{i+1},\dots,a_k],\\
g|^{[k]}&=[a_1,a_2,\dots,a_{i+1},a_{i},\dots,a_k].
\end{align*}
for some $i\in[k-1]$, i.e., $f|^{[k]}$ and $g|^{[k]}$ differ by an
adjacent transposition of the $i$th and $(i+1)$st elements.

Let us now define $\Delta=a_{i+1}-a_i$. It follows that
\[\Phi(f)_{k+1}-\Phi(g)_{k+1} \equiv 2\Delta \pmod{k}.\]
Since $1\leq \abs{\Delta} \leq k-1$ and $k\geq 3$ is a prime, we know
that $2\Delta$ is not a multiple of $k$. As a result, we get
\[\abs{\Phi(f)_{k+1}-\Phi(g)_{k+1}}\geq  1.\]

Similarly, we have
\begin{align*}
\Phi(f)_{k+2}-\Phi(g)_{k+2} &\equiv(2i-1)^2 a_i+ (2i+1)^2 (a_i+\Delta)\\
& \quad\ -(2i-1)^2(a_i+\Delta)-(2i+1)^2 a_i\\
&\equiv 8i\Delta \pmod{k}.
\end{align*}
Again, $8i\Delta$ is not a multiple of $k$ since $1\leq i,
\abs{\Delta}\leq k-1$ and $k\geq 3$ is a prime. This implies that
\[\abs{\Phi(f)_{k+2}-\Phi(g)_{k+2}}\geq  1.\]

Thus, by Lemma \ref{theorem1}, we get
\begin{align*}
d_K(f,g)&\geq d_K(f|^{[k]},g|^{[k]})+\sum_{i=k+1}^{k+2}\abs{\Phi(f)_i-\Phi(g)_i}\\
&\geq 1+1+1 = 3.
\end{align*}

\paragraph{Case 2} $d_K(f|^{[k]},g|^{[k]})=2$. Let us denote
\[f|^{[k]}=[a_1,a_2,\dots,a_k].\]
By our assumption, there exist $1\leq i,j\leq k-1$ such $g$ is
obtained from $f$ as a result of two adjacent transpositions: one
exchanging locations $i$ and $i+1$, and one exchanging locations $j$
and $j+1$. We distinguish between two cases.

In the first case, $\mathset{i,i+1}\cap\mathset{j,j+1}=\emptyset$. Without loss
of generality, assume $i<j$, and then we have
\[g|^{[k]}=[a_{1},...,a_{i+1},a_{i},...,a_{j+1},a_{j},...,a_{k}].\]
Let us define $\Delta_1=a_{i+1}-a_i$, and $\Delta_2=a_{j+1}-a_j$. Then we get
\[\Phi(f)_{k+1}-\Phi(g)_{k+1} \equiv 2(\Delta_1+\Delta_2)\pmod{k}.\]
If $\Delta_1+\Delta_2$ is not a multiple of $k$, then by the same reasoning as
before, 
\[\abs{\Phi(f)_{k+1}-\Phi(g)_{k+1}}\geq 1.\]
If $\Delta_1+\Delta_2$ is a multiple of $k$, then we can write
$\Delta_2\equiv-\Delta_1\pmod{k}$. Hence,
\begin{align*}
\Phi(f)_{k+2}-\Phi(g)_{k+2} &\equiv (2i-1)^2a_i+ (2i+1)^2 (a_i+\Delta_1)\\
&\quad \ +(2j-1)^2a_j+(2j+1)^2 (a_j-\Delta_1)\\
&\quad \ -(2i-1)^2(a_i+\Delta_1)-(2i+1)^2 a_i\\
&\quad \ -(2j-1)^2(a_j-\Delta_1)-(2j+1)^2 a_j\\
&\equiv 8(j-i)\Delta_1 \pmod{k}.
\end{align*}
Since $8(j-i)\Delta_1$ is not a multiple of $k$, we have
\[\abs{\Phi(f)_{k+2}-\Phi(g)_{k+2}}\geq 1.\]

In the second case,
$\mathset{i,i+1}\cap\mathset{j,j+1}\neq\emptyset$. Thus, either
\[g|^{[k]}=[a_1,...,a_{i+2},a_i,a_{i+1},...,a_k],\]
or
\[g|^{[k]}=[a_1,...,a_{i+1},a_{i+2},a_i,...,a_k],\]
for some $i\in[k-2]$. By defining $\Delta_1=a_{i+2}-a_{i+1}$ and $\Delta_2=a_{i+2}-a_i$ in the first case, or $\Delta_1=a_{i+1}-a_i$ and $\Delta_2=a_{i+2}-a_i$ in the second case, and with the same arguments as above, it can be proved that either
\[\abs{\Phi(f)_{k+1}-\Phi(g)_{k+1}}\geq 1,\]
or
\[\abs{\Phi(f)_{k+2}-\Phi(g)_{k+2}}\geq 1.\]
Combining all the cases together, by Lemma \ref{theorem1}, we get
\begin{align*}
d_K(f,g)&\geq d_K(f|^{[k]},g|^{[k]})+\sum_{i=k+1}^{k+2}\abs{\Phi(f)_i-\Phi(g)_i}\\
&\geq 2+1 = 3.
\end{align*}

\paragraph{Case 3} $d_K(f|^{[k]},g|^{[k]})\geq 3$. This is the easiest case, since
by Lemma \ref{theorem1},
\[d_K(f,g) \geq d_K(f|^{[k]},g|^{[k]}) \geq 3.\]

Finally, we note that if $k+1$ is a prime, we can repeat the proof in
its entirety, replacing $\bmod k$ with $\bmod (k+1)$.
\end{IEEEproof}

Before continuing to the next construction we would like to consider
encoding and decoding algorithms for the code from Construction
\ref{constr:1}. For the encoding procedure, we start by mapping an
integer from $\Z_{k!}$ to a permutation $f'\in S_k$. This may be
accomplished in linear time \cite{MarStr07}. Then, using the
description of Construction \ref{constr:1}, the two redundancy symbols
are easily placed in their correct position, and we receive a codeword
$f\in C$ such that $f|^{[k]}=f'$.

Decoding may be done efficiently as well. Assume $f\in C\subseteq
S_{k+2}$ was transmitted, while $g\in S_{k+2}$ was received, where
$d_K(f,g)\leq 1$.  A trivial decoding algorithm can check the $k+2$
permutation in the ball of radius $1$ centered around $g$, and find
the unique codeword $f$ in it. This algorithm takes $O(k^2)$ steps.

We can do better than that, using the decoding algorithm we now
describe. Let $\hat{g}\in C$ be the unique codeword in $C$ having the
same order of information symbols as $g$, i.e.,
$\hat{g}|^{[k]}=g|^{[k]}$. If $d_K(\hat{g},g)\leq 1$, then $f=\hat{g}$
is the correct decoding. Otherwise, $d_K(f|^{k},g|^{k})=1$, and we can
write
\begin{align*}
f|^{[k]}&=[a_1,\dots,a_i,a_{i+1},\dots,a_k],\\
g|^{[k]}&=[a_1,\dots,a_{i+1},a_{i},\dots,a_k],
\end{align*}
for some $i\in [k-1]$.

Since a single adjacent transposition changed the order of two information symbols, we deduce no redundancy symbols were moved, and thus,
\[ \Phi(f)_{k+1}=\Phi(g)_{k+1} \qquad \text{and} \qquad
\Phi(f)_{k+2}=\Phi(g)_{k+2}.\]
According to the proof of Theorem \ref{thm:oneCode},
\begin{align*}
\Phi(g)_{k+1} - \Phi(\hat{g})_{k+1} &\equiv 2(a_{i+1}-a_i) \pmod{m}, \\
\Phi(g)_{k+2} - \Phi(\hat{g})_{k+2} &\equiv 8i(a_{i+1}-a_i) \pmod{m},
\end{align*}
where $m$ is the prime in $\mathset{k,k+1}$. Combining the two
equations together we get
\[\Phi(g)_{k+2} - \Phi(\hat{g})_{k+2} \equiv 4i \parenv{\Phi(g)_{k+1} - \Phi(\hat{g})_{k+1}} \!\!\!\pmod{m},\]
and we can easily solve for $i$, thus recovering the coordinate of the
adjacent transposition. This decoding algorithm runs in $O(k)$ steps.
We illustrate the decoding algorithm with the following example.

\begin{example}
Let $k=4$, and assume we would like to encode $[4,1,3,2]$. Thus, by
Construction \ref{constr:1}, we look for a permutation $f\in S_6$ such
that
\begin{align*}
\Phi(f)_5 & = \parenv{\sum_{i=1}^{4}(2i-1) f(i)}\bmod 5 = 1,\\
\Phi(f)_6 & = \parenv{\sum_{i=1}^{4}(2i-1)^2 f(i)}\bmod 5 = 1.
\end{align*}
We therefore transmit
\[f=[4,1,3,5,6,2],\]
and let us assume the received permutation is
\[g=[4,3,1,5,6,2],\]
due to an adjacent transposition in positions $2$ and $3$. We extract
the information symbols from $g$ to obtain,
\[g|^{[k]}=[4,3,1,2],\]
and use that to construct the codeword
\[\hat{g}=[4,6,3,5,1,2].\]
Since $d_K(\hat{g},g)>1$, we deduce that two information symbols
changed positions. Since
\begin{align*}
\Phi(g)_{k+1} & = 1 & \Phi(g)_{k+2} & = 1 \\
\Phi(\hat{g})_{k+1} & = 2 & \Phi(\hat{g})_{k+2} & = 4,
\end{align*}
we solve
\[ 1-4 \equiv 4i(1-2) \pmod{5},\]
resulting in the correct positions of the adjacent transposition, $i=2$
and $i+1=3$.
\end{example}

Another strategy for constructing rank-modulation codes for Kendall's
$\tau$-metric, which was already employed by
\cite{JiaSchBru10,BarMaz10}, is to first construct a code $C^*$ with
minimum $\ell_1$-distance $d$ in $\Z^n$, and then take
\[C=\Phi^{-1}(C^*\cap\Z_n!),\]
i.e., exactly those permutations whose factoradic representations are
$C^*\cap\Z_n!$. Since by \eqref{eq:distl1}, the distance can only
increase, the resulting set of permutations is a code with minimum
Kendall's $\tau$-distance of at least $d$. The main challenge with
this approach is to ensure a large intersection of $C^*$ with $\Z_n!$.

For the construction of systematic codes we shall employ the same
methods, however, now we have an additional
challenge: we also require the intersection $C^*\cap\Z_n!$ to have at
least one vector for each possible prefix from $\Z_k!$.

\begin{construction}
\label{con:1ecc}
Let $k\geq 2$ be some integer. For a vector
$x=(x_1,x_2,\dots,x_{k+1})\in\Z^{k+1}$ we denote
\[s_m(x)=\parenv{\sum_{i=1}^m ix_i}\bmod (2k+3).\]
We construct a subset $C'\subseteq\Z^{k+1}$ defined by
\begin{align*}
C'=\{x\in\Z^{k+1} ~|~  &x_{k}=\floorenv{s_{k-1}(2x)/3},\\
 &x_{k+1}=s_{k-1}(2x) \bmod 3\}.
\end{align*}
We denote by
\[C^* = \mathset{ (0,x_1,x_2,\dots,x_{k+1}) ~|~ (x_1,x_2,\dots,x_{k+1})\in C'},\]
the prepending of $0$ to all the codewords of $C'$. The constructed
code is
\[C=\Phi^{-1}\parenv{C^*\cap\Z_{k+2}!}.\]
\end{construction}

\begin{theorem}
For all $k\geq 2$, the code $C$ from Construction \ref{con:1ecc} is a
$[k+2,k,3]$ systematic code in Kendall's $\tau$-metric.
\end{theorem}

\begin{IEEEproof}
Consider the perfect $(k+1)$-dimensional single-error-correcting code
in the $\ell_1$-metric described by Golomb and Welch in
\cite{GolWel70} and given by,
\[C''=\mathset{\left. x=(x_1,x_2,\dots,x_{k+1})\in\Z^{k+1} ~\right|~ s_{k+1}(x)=0 }.\]
We contend that $C'\subseteq C''$, i.e., that $C'$ is also a
single-error-correcting code in the $\ell_1$-metric. Indeed, let
$x=(x_1,\dots,x_{k+1})\in C'$ be a codeword in $C'$. Then, noting
that
\[k\equiv 3(k+1) \pmod{2k+3},\]
and working modulo $2k+3$, and we get
\begin{align*}
s_{k+1}(x) & \equiv s_{k-1}(x) + kx_{k} + (k+1)x_{k+1}\\
& \equiv s_{k-1}(x) + k\floorenv{s_{k-1}(2x)/3} \\
& \quad \ + (k+1)\parenv{s_{k-1}(2x) \bmod 3} \\
& \equiv s_{k-1}(x) + (k+1) (3\floorenv{s_{k-1}(2x)/3} \\
& \qquad \qquad \qquad \qquad \qquad + (s_{k-1}(2x) \bmod 3))\\
& \equiv s_{k-1}(x) + (k+1)s_{k-1}(2x) \\
& \equiv s_{k-1}(x) + 2(k+1) s_{k-1}(x) \\
& \equiv (2k+3)s_{k-1}(x) \equiv 0 \pmod{2k+3}.
\end{align*}
Thus, $x\in C''$, and so $C'\subseteq C''$.

We note how the first $k-1$ coordinates of the codewords of $C'$ are
unconstrained. Thus, for all $1\leq i\leq k-1$ we can set
$x_{i}\in\Z_{i+1}$ arbitrarily in any one of $k!$
ways. Furthermore, for any $x\in C'$,
\[ 0\leq \floorenv{s_{k-1}(2x)/3} \leq \frac{2(k+1)}{3} \leq k,\]
as well as
\[ 0\leq s_{k-1}(2x) \bmod 3 \leq k+1.\]
It follows that $x_{k}\in \Z_{k+1}$ and $x_{k+1}\in \Z_{k+2}$. Hence, after
prepending a $0$ to the codewords to obtain $C^*$, we get
\[\abs{C^*\cap \Z_{k+2}!} = k!.\]
Finally, prepending the $0$ does not change the minimum distance, and so
$C^*$ has minimum $\ell_1$-distance of $3$, and therefore, so does
the final constructed code $C=\Phi^{-1}(C^*\cap\Z_{k+2}!)$.
\end{IEEEproof}

Encoding the code from Construction \ref{con:1ecc} is extremely
easy. In the factoradic representation we arbitrarily fill in the
first $k-1$ entries. The last two digits are determined by the first
$k-1$ digits, and a $0$ is then prepended. We then convert the
factoradic representation to a permutation, which is the desired
codeword. The entire procedure takes $O(k)$ steps if we use
\cite{MarStr07} to convert from the factoradic representation to
permutations.

The decoding process is simple as well. Given a permutation read from
the channel, we first translate it to its factoradic representation
and remove the leading $0$. The remaining $k+1$ coordinates are
decoded using any simple procedure for decoding the Golomb-Welch code
\cite{GolWel70}. Again, the entire procedure takes $O(n)$ steps.

We note that the two constructions are not equivalent. As an example,
the $[5,3,3]$ code from Construction \ref{constr:1} contains the codewords
\[f=[1,4,3,2,5] \quad \text{and} \quad g=[2,3,4,1,5].\]
However,
\[\Phi(f)=[0,0,1,2,0] \quad \text{and} \quad \Phi(g)=[0,1,1,1,0].\]
Since
\[d_1(\Phi(f),\Phi(g)) = 2,\]
the code cannot have originated from Construction \ref{con:1ecc}.

An obvious question to ask is how good are the parameters of the codes
presented in Construction \ref{constr:1} and Construction
\ref{con:1ecc}. Any $(n,M,d)$ code (systematic or not) has to satisfy
the ball-packing bound:
\begin{equation}
\label{eq:perfect}
M \leq \frac{n!}{\abs{\mathfrak{B}_r(n)}},
\end{equation}
where $r=\floorenv{(d-1)/2}$. Codes attaining \eqref{eq:perfect} with
equality are called \emph{perfect}. We thus reach the following simple
corollary:
\begin{corollary}
The $[k+2,k,3]$ systematic codes of Construction \ref{constr:1} and
Construction \ref{con:1ecc} have optimal size, unless perfect
systematic one-error-correcting codes exist in Kendall's
$\tau$-metric.
\end{corollary}

Example of perfect codes in other metrics are quite rare (see
\cite{MacSlo78}). In Kendall's $\tau$-metric there is a simple
$(3,2,3)$ that is perfect:
\[C=\mathset{[1,2,3],[3,2,1]}.\]
This code is also systematic, i.e., a $[3,2,3]$-code. However, beside
this code, no other perfect code has been found yet. It was recently
shown in \cite{BuzEtz13}, that no perfect codes exist in $S_n$ under
Kendall's $\tau$-metric when $n$ is a prime, or when $4\leq n\leq 10$.

To summarize, the $[k+2,k,3]$ codes presented have minimal redundancy
among systematic codes, unless there exists a perfect systematic
$[k+1,k,3]$ one-error-correcting code. Furthermore, compared with the
one-error-correcting code presented in \cite{JiaSchBru10}, the codes
presented here have more efficient encoding and decoding algorithms.

\subsection{Multi-Error-Correcting Codes}
\label{sec:ktmulti}

After studying systematic one-error-correcting codes, we turn to
consider systematic codes capable of correct more than one error. We
will first describe an explicit construction for a wide range of
parameters, and then turn to a greedy algorithm leading us to prove a
non-constructive existence result.

The systematic one-error-correcting code in Construction
\ref{constr:1} may be generalized in a straightforward way: for $1\leq
k\leq n$ and $r\geq 1$ integers we define,
\[C = \mathset{f\in S_{k+r} ~\left|~ \text{$\Phi(f)_{k+j}=\rho_j(f|^{[k]})$ for
all $j\in[r]$}\right.},\]
where $\rho_j(\cdot)$ is given by \eqref{eq:rho}. This gives us a
family of codes, including a $[10,4,5]$ systematic code, and a
$[14,4,7]$ systematic code. However, a general analysis of these codes
is difficult.

We therefore return to the strategy of metric embedding: an
$\ell_1$-metric code is constructed in such a way as to allow all
possible values in the first few information entries, and then the
other positions are determined as a function of the information
entries.

\begin{construction}
\label{con:genecc}
Let $p$ be a prime, $m\geq 2$ an integer, $1\leq t\leq
\frac{p-3}{2}$ also an integer, and
\[\max(p^{m-1},p+tm-1)\leq n\leq p^m-1.\]
Arbitrarily choose $\alpha_1,\alpha_2,\dots,\alpha_n$ to be $n$
distinct non-zero elements of $\gf(p^m)$. We define
\[H=\begin{bmatrix}
1 & 1 & \dots & 1 \\
\alpha_1 & \alpha_2 & \dots & \alpha_n \\
\alpha_1^2 & \alpha_2^2 & \dots & \alpha_n^2 \\
\vdots & \vdots & \dots & \vdots \\
\alpha_1^{t} & \alpha_2^{t} & \dots & \alpha_n^{t}
\end{bmatrix}.\]
Viewing $\gf(p^m)$ as the vector space $\gf(p)^m$, we can think of any
entry of the form $\alpha_i^j$ in $H$ as a column vector of length $m$
over $\gf(p)$. Thus, we shall consider $H$ to be a $(t+1)m\times n$ matrix
over $\gf(p)$. We denote
\[k = n-\rank H.\]

We construct a subset $C'\subseteq\Z^n$ defined by
\[C'=\mathset{x\in\gf(p)^n ~|~ Hx\equiv 0 \pmod{p} },\]
where the entries of $Hx$ are computed modulo $p$.

We define the mapping $\mu:\Z^n\to\Z^{n+1}$ as follows,
\begin{multline*}
\mu(x_1,x_2,\dots,x_n) =  \\
= (0, x_1,\dots,x_k, x_{k+1}\bmod p, \dots, x_{n}\bmod p),
\end{multline*}
i.e., prepending a zero and reducing the last $n-k$ entries modulo
$p$. We then set
\[C^* = \mathset{ \mu(x) ~|~ x\in C'}.\]
The constructed code is
\[C = \Phi^{-1}(C^* \cap \Z_{n+1}!).\]
\end{construction}

\begin{theorem}
The code $C$ from Construction \ref{con:genecc} is an $[n+1,k+1,2t+2]$
systematic code in Kendall's $\tau$-metric capable of correcting $t$
errors. Furthermore, the code's redundancy satisfies $n-k\leq tm+1$.
\end{theorem}

\begin{IEEEproof}
The matrix $H$ is nothing but the parity-check matrix for a BCH code
over $\gf(p)$. Since the code is linear, we can find a $k\times n$
generator matrix $G$ for the code, and in particular, we can require
that it be systematic, i.e.,
\[G = [ I_k | A ],\]
where $I_k$ is the $k\times k$ identity matrix, and $A$ is a $k\times
(n-k)$ matrix over $\gf(p)$. As a side note, getting to this
systematic form, we may be required to permute the coordinates of the
code. Since the order of elements $\alpha_1,\dots,\alpha_n$, which are
used to construct $H$, is irrelevant, we assume it is chosen so that
no change of order of coordinates is required.

Let us denote by $C''$ the code whose generator matrix is $G$. We
recall the definition of the Lee-distance measure over $\gf(p)$: given
two vectors $x,x'\in \gf(p)^n$,
\[d_L(x,x') = \sum_{i=1}^n \min\parenv{\abs{x_i-x'_i},\abs{x'_i-x_i}},\]
where subtraction is done in $\gf(p)$. It was shown in
\cite{RotSie94}, that the minimum Lee distance of the code $C''$ is at
least $2t+2$.

Our next goal is to transform $C''$ to a code over $\Z^n$ with a
minimum $\ell_1$-distance guarantee. Since we have the code $C''$
reside in the $n$-dimensional cube $\gf(p)^n$, we can place copies of
that cube and tile the entire space $\Z^n$. This is known as
Construction A of \cite{LeeSlo71}, and the resulting code is exactly
\[C'=\mathset{x\in\gf(p)^n ~|~ Hx\equiv 0 \pmod{p} },\]
where the entries of $Hx$ are computed modulo $p$. Again, by
\cite{LeeSlo71}, the codewords of $C'$ are spanned (using linear
combinations with integer coefficients) by the generating matrix
\[G'=\begin{bmatrix}
I_k & A \\
0 & pI_{n-k}
\end{bmatrix}.
\]
Thus, the minimum Lee distance of $2t+2$ between codewords of $C''$
guarantees a minimum $\ell_1$-distance of $2t+2$ between codewords of
$C'$.

A quick inspection of $G'$ reveals that, due to the first $k$ rows,
any prefix of $k$ integers may be completed to a length-$n$ codeword
in $C'$. Furthermore, given a codeword in $C'$, by reducing its last
$n-k$ entries modulo $p$ we obtain another codeword of $C'$, due to
the last $n-k$ rows of $G'$. It follows that
\[C^* = \mathset{ \mu(x) ~|~ x\in C'},\]
is a subset of the codewords of $C'$ with a $0$ prepended, and that
\[\abs{C^*\cap\Z_{n+1}!} = (k+1)!.\]
Hence, $C$ is indeed an $[n+1,k+1,2t+2]$ systematic code in Kendall's
$\tau$-metric.

Finally, it is well known (see also \cite{RotSie94}) that when $H$ is
viewed as a $(t+1)m\times n$ matrix over $\gf(p)$,
\[n-k = \rank H \leq tm+1.\]
\end{IEEEproof}

Again, encoding and decoding are easily done. For an encoding
procedure, take any vector $(0|u)\in \Z_{k+1}!$ and map it to
\[ (0|u) \mapsto (0 | u | uA \bmod p) \in C^*.\]
The permutation whose factoradic representation is given by this
vector is the encoded permutation.

For a decoding procedure, map the received permutation to its
factoradic representation, and use the decoding for the Lee-metric
code (essentially, a BCH decoding procedure) given in \cite{RotSie94}.

We also note that for the least redundancy, we would like to choose
$m=2$ in Construction \ref{con:genecc}. Using Bertrand's postulate,
that any interval $[s,2s]$ contains a prime, $s$ a positive integer,
we can show the following corollary:

\begin{corollary}
For any $t\geq 1$, and $n\geq 6t+5$, there exists a prime $p$, such
that the requirements of Construction \ref{con:genecc} are satisfied
with $m=2$, and therefore there exists an $[n+1,k+1,2t+2]$ systematic
code with redundancy at most $2t+1$.
\end{corollary}

Along the same lines, but using two embeddings, one after the other,
we present a construction transforming systematic binary codes under
the Hamming metric, into systematic codes of permutations under
Kendall's $\tau$-metric. The construction is a simple modification of
the construction given in \cite{MazBarZem11}.

The main idea for the first embedding is to use a mapping $\cG_m:
\Z_{2^m}\rightarrow \mathset{0,1}^m$ such that for any two integers
$t_1,t_2\in \Z_{2^m}$,
\begin{equation}
\label{eq:gray}
\abs{t_1-t_2} \geq d_H(\cG_m(t_1),\cG_m(t_2)),
\end{equation}
where $d_H(\cdot,\cdot)$ denotes the Hamming distance function. By
convention, $\cG_0$ is the mapping returning the unique vector of
length $0$. A simple way of creating such a mapping is to use the
encoding function for an optimal Gray code (see \cite{Sav97} for a
survey of Gray codes).

\begin{construction}
\label{con:twoembed}
Let $C'$ be an $(n',2^{k'},d)$ binary systematic code in the Hamming
metric, where the first $k'$ coordinates are systematic. Furthermore,
let $k$ and $n$ be integers such that
\begin{align*}
k' &= \sum_{i=1}^k \ceilenv{\log_2 i},  \\
n' &= \sum_{i=1}^k \ceilenv{\log_2 i} + \sum_{i=k+1}^{n}\floorenv{\log_2 i}.
\end{align*}
We conveniently define
\[\lambda(i) = \begin{cases}
\ceilenv{\log_2 i} & 1\leq i\leq k, \\
\floorenv{\log_2 i} & k+1 \leq i\leq n.
\end{cases}\]
We now construct the following code,
\[
C = \mathset{ f\in S_n ~\left|~ \cG_{\lambda(1)}(\Phi(f)_1)||\dots ||\cG_{\lambda(n)}(\Phi(f)_n)\in C'\right.}.
\]
where $||$ denotes vector concatenation. In particular, the notation implies
that for any $f\in C$, 
\[ 0 \leq \Phi(f)_i \leq 2^{\floorenv{\log_2 i}}-1\leq i-1,\]
for all $k+1\leq i\leq n$.
\end{construction}

\begin{theorem}
The code $C$ from Construction \ref{con:twoembed} is an $[n,k,d]$
systematic code in Kendall's $\tau$-metric.
\end{theorem}

\begin{IEEEproof}
The length of the code is obviously $n$. Let us try to build a
codeword $f\in C$. We note that the first $k$ symbols of $\Phi(f)$
form a binary vector of length $k'$ after being Gray-mapped and
concatenated. Since the first $k'$ bits of the code $C'$ are
systematic, any such $k'$-prefix may be uniquely completed to form a
codeword in $C'$ by adding appropriate $n'-k'$ redundancy bits. These
redundancy bits can be divided into sets of size $\floorenv{\log_2
  i}$, with $k+1\leq i\leq n$. Thus, the reverse Gray mapping of these
sets uniquely determines $\Phi(f)_{k+1},\dots,\Phi(f)_{n}$, and
therefore, $f$ as well. It follows that $C$ is indeed a systematic
code of length $n$ and $k$ information symbols.

Finally, let $f,g\in C$ be two distinct codewords. Then, using
\eqref{eq:distl1} and \eqref{eq:gray} we get
\begin{align*}
d_K(f,g) & \geq \sum_{i=1}^n \abs{\Phi(f)_i-\Phi(g)_i} \\
& \geq \sum_{i=1}^n d_H\parenv{\cG_{\lambda(i)}(\Phi(f)_i),\cG_{\lambda(i)}(\Phi(g)_i)}\\
& \geq d.
\end{align*}
Thus, $C$ is an $[n,k,d]$ systematic code.
\end{IEEEproof}

We turn to present a generic scheme for constructing an $[n,k,d]$
systematic codes. The scheme is a simple adaptation of the
Gilbert-Varshamov lower bound. Although it is beyond the scope of this
paper to obtain efficient encoding and decoding algorithms for it, the
analysis of this scheme is very useful for proving the existence of
codes with certain parameters, and for deriving the capacity of
systematic codes.

\begin{construction}\label{constr:2}
Let $2\leq k < n$ and $d\geq 1$ be integers. We define $C_0=\emptyset$.
For all $i\geq 1$, the algorithm searches for $f\in S_n$ such that
\begin{equation}
\label{eq:req1}
\min_{g\in C_{i-1}}d_K(f,g) \geq d,
\end{equation}
and
\begin{equation}
\label{eq:req2}
f|^{[k]} \not\in \mathset{\left. g|^{[k]} ~\right|~ g\in C_{i-1}}.
\end{equation}
If we can continue the process, increasing $i$ by $1$ at each
iteration, and reach $i=k!$, then the constructed code is
$C=C_{k!}$. Otherwise, we declare failure.
\end{construction}

\begin{theorem}
If the algorithm of Construction \ref{constr:2} succeeds, the
resulting code $C$ is a systematic $[n,k,d]$ code in Kendall's
$\tau$-metric.
\end{theorem}
\begin{IEEEproof}
The proof is straightforward. We start with an empty set, and at each stage
we add a codeword that is not to close to previously chosen codewords. Thus,
the minimum distance of the resulting code is $d$. The second requirement
at each step, is that the information symbols do not repeat those of a
previously chosen codeword. Thus,
\[\mathset{\left. f|^{[k]} ~\right|~ f\in C}=S_k,\]
and the code is systematic.
\end{IEEEproof}

For Construction \ref{constr:2} to succeed, the number of redundancy
symbols, $n-k$, needs to be sufficiently large. In the following
theorem, we derive a bound for these parameters.

\begin{theorem}
\label{theorem2}
Construction \ref{constr:2} can successfully build an $[n,k,d]$
systematic code if
\[\sum_{i=1}^{d-1}\binom{k+i-2}{i}\binom{d-i-1+n-k}{n-k}2^{\min{(d-i-1,n-k)}} <\frac{n!}{k!}.\]
\end{theorem}

\begin{IEEEproof}
For any permutations $h\in S_k$ there are exactly $n!/k!$ permutations
$f\in S_n$ such that $f|^{[k]}=h$. At each step $i$ of the algorithm
we shall arbitrarily choose $h\in S_k$ such that
\[h \not\in \mathset{\left. g|^{[k]} ~\right|~ g\in C_{i-1}}.\]
We shall then try to find $f\in S_n$ such that $f|^{[k]}=h$, i.e., $f$
satisfies requirement \eqref{eq:req2}. Our goal is to show there is at
least one such $f$ that also satisfies the requirement of
\eqref{eq:req1}.

Given any such $h\in S_k$, let us upper bound the number of
permutations $f$ such that $f|^{[k]}=h$ but $f$ does not satisfy
\eqref{eq:req1}. Let $g\in C_{i-1}$ be a codeword chosen in some
previous iteration, and assume $d_K(f,g)\leq d-1$. Let us denote
\[d_K(f|^{[k]},g|^{[k]}) = j \leq  d-1.\]
By Lemma \ref{theorem1}, in order for us to have $d_K(f,g)\leq d-1$,
we must have
\[\sum_{t=1}^{n-k}\abs{\Phi(f)_{k+t}-\Phi(g)_{k+t}} \leq d-j-1.\]
Thus, we would like to count the number of integer vectors of length
$n-k$, whose $\ell_1$ weight is at most $d-j-1$. Choosing the
magnitudes of the entries of such a vector is equivalent to the number
of ways $d-j-1$ identical balls can be placed in $n-k+1$ non-identical
bins.  We also need to choose the sign for the non-zero entries of
such a vector, and there are at most $\min(d-j-1,n-k)$ such entries.
It follows, that an upper bound on the number of permutations $f\in
S_n$ such that $f|^{[k]}=h$, and $d_K(f|^{[k]},g|^{[k]})=j$ for the
given $g$, is
\[\binom{d-j-1+n-k}{n-k}2^{\min(d-j-1,n-k)}.\]

Let $N_j$ denote the number of permutations $g\in C_{i-1}$ such that
$d_K(f|^{[k]},g|^{[k]})=j$. If we had this number, then by a simple
union bound, the total number of permutations $f\in S_n$ such that
$f|^{[k]}=h$, but \eqref{eq:req1} does not hold, is upper bounded by
\[\sum_{j=1}^{d-1}N_j \binom{d-j-1+n-k}{n-k}2^{\min{(d-j-1,n-k)}}.\]

To continue our upper bound, we replace $N_j$ with the larger $N'_j$,
where $N'_j$ denotes the number permutations $h'\in S_k$ such that
$d_K(h',h)=j$. Our upper bound is now
\begin{equation}
\label{eq:upb}
\sum_{j=1}^{d-1}N'_j \binom{d-j-1+n-k}{n-k}2^{\min{(d-j-1,n-k)}}.
\end{equation}

According to Lemma \ref{lemma2},
\[\sum_{t=0}^j N'_t = \abs{\mathfrak{B}_j(k)} \leq \binom{k+j-1}{k-1}.\]
In this case, it is not hard to prove that \eqref{eq:upb} is maximized
when
\[N'_j=\binom{k+j-1}{k-1}-\binom{k+j-2}{k-1}=\binom{k+j-2}{k-2},\]
for $k\geq 2$ and $1\leq j\leq d-1$, since
\[F(j)=2^{\min{(d-j-1,n-k)}}\binom{d-j-1+n-k}{n-k},\]
is a deceasing function in $j$.

As a result, the number of permutations $f\in S_n$, such that
$f|^{[k]}=h$, but \eqref{eq:req1} does not hold, is upper bounded by
\begin{equation}
\label{eq:upb1}
\sum_{i=1}^{d-1}\binom{k+i-2}{i}\binom{d-i-1+n-k}{n-k}2^{\min{(d-i-1,n-k)}}.
\end{equation}
Since the total number of permutations $f\in S_n$ such that
$f|^{[k]}=h$ is $n!/k!$, if \eqref{eq:upb1} is strictly less than
$n!/k!$ then there exists a permutation $f$ satisfying
\eqref{eq:req1}. Since we did not restrict $h$ in any way, this
conclusion holds for any iteration of the algorithm, and the algorithm
succeeds.
\end{IEEEproof}

\begin{example}
When $d=3$ and $n=k+2$, the inequality of Theorem \ref{theorem2} can
be simplified to
\[6\binom{k-1}{1} +\binom{k}{2}< (k+1)(k+2),\]
which holds for any $k\geq 2$. Therefore, there exists a $[k+2,k,3]$
systematic code for any $k\geq 2$. Note that this result is consistent
with the codes built in Construction \ref{constr:1} and Construction
\ref{con:1ecc}.
\end{example}

\begin{example}
When $d=4$ and $n=k+3$, the inequality of Theorem \ref{theorem2} can
be simplified to
\[40\binom{k-1}{1} +8\binom{k}{2} +\binom{k+1}{3}< (k+1)(k+2)(k+3),\]
which holds for all $k\geq 2$. Therefore, there exists a $[k+3,k,4]$
systematic code for any $k\geq 2$.
\end{example}

We now prove our main existential result, which is non-constructive in
nature, since it builds upon Construction \ref{constr:2}.

\begin{theorem}
\label{theorem3}
There exists a $[k+d,k,d]$ systematic code in Kendall's $\tau$-metric,
for any $k\geq 2$ and $d\geq 1$.
\end{theorem}
\begin{IEEEproof}
Based on Theorem \ref{theorem2}, to show that there exists a $[k+d,k,d]$
systematic code, we only need to prove
\[\sum_{i=1}^{d-1}\binom{k+i-2}{i}\binom{2d-1-i}{d}2^{d-i-1}<\frac{(k+d)!}{k!}\]
for $k\geq 2$ and $d\geq 1$. We note that the case $d=1$ is trivial,
and so we will assume throughout the rest of the proof that $d\geq 2$.
Furthermore, to simplify the proof, we will prove a stronger claim,
\begin{equation}
\label{equ2_6}
\sum_{i=1}^{d-1}\binom{k+i}{i}\binom{2d-1-i}{d}2^{d-i-1}<\frac{(k+d)!}{k!}.
\end{equation}

Let us define
\[\psi_d(k)=\frac{k!}{(k+d)!}\sum_{i=1}^{d-1}\binom{k+i}{i}\binom{2d-1-i}{d}2^{d-i-1}.\]
We contend the $\psi_d(k)$ is non-increasing in $k$, and to prove this claim we
consider $\psi_d(k+1)/\psi_d(k)$ and note that
\begin{align*}
\frac{\psi_d(k+1)}{\psi_d(k)}&=
\frac{\binom{k}{d}}{\binom{k+1}{d}}\cdot
\frac{\sum_{i=1}^{d-1}\binom{k+1+i}{i}\binom{2d-1-i}{d}2^{d-i-1}}{\sum_{i=1}^{d-1}\binom{k+i}{i}\binom{2d-1-i}{d}2^{d-i-1}}\\\
&=\frac{\binom{k}{d}}{\binom{k+1}{d}}\cdot
\frac{\sum_{i=1}^{d-1}\frac{k+1+i}{k+1}\binom{k+i}{i}\binom{2d-1-i}{d}2^{d-i-1}}{\sum_{i=1}^{d-1}\binom{k+i}{i}\binom{2d-1-i}{d}2^{d-i-1}}\\
	&\leq \frac{k+1}{k+d+1} \cdot\frac{k+d}{k+1}< 1.
\end{align*}
Thus, $\psi_d(k)$ is indeed a non-increasing function of $k$. If
$\psi_d(2)<1$ for all $d\geq 2$, then for any $k, d\geq 2$, we surely
have $\psi_d(k)< 1$, which proves \eqref{equ2_6}. So our task is to
prove $\psi_d(2)<1$, namely,
\begin{equation}
\label{eq:toshow}
\sum_{i=1}^{d-1}\binom{2+i}{i}\binom{2d-1-i}{d}2^{d-i-1}< \frac{(2+d)!}{2!},
\end{equation}
for all $d\geq 2$.

For all $2\leq d\leq 16$ we can show that the inequality holds by
computing the exact values.  In what follows, we show that the
inequality also holds when $d>16$. The left-hand side of \eqref{eq:toshow}
may be upper bounded by
\begin{multline*}
\sum_{i=1}^{d-1}\binom{2+i}{i}\binom{2d-1-i}{d}2^{d-i-1}\leq \\
\leq \frac{d(d+1)(d+2)}{2}\binom{2d-2}{d}2^{d-2}.
\end{multline*}
Thus, to prove \eqref{eq:toshow}, it suffices to prove
\[\binom{2d-2}{d}2^{d-2} < (d-1)!.\]
We define
\[\xi(d)=\frac{1}{(d-1)!}\binom{2d-2}{d}2^{d-2}.\]
We can numerically check that $\xi(17)<1$, and since
\[\frac{\xi(d+1)}{\xi(d)} = \frac{4d(2d+1)}{d(d+1)(d-1)}<1\]
for all $d>16$, we have $\xi(d)<1$ for all $d>16$, and this completes
the proof.
\end{IEEEproof}

\subsection{Capacity of Systematic Codes}
\label{sec:ktcap}

In this section, we prove that for rank modulation under Kendall's
$\tau$-metric, systematic error-correcting codes achieve the same
capacity as general error-correcting codes.

In \cite{BarMaz10}, Barg and Mazumdar derived the capacity of general
error-correcting codes for rank modulation under Kendall's
$\tau$-metric. Let $A(n,d)$ denote the maximum size of an $(n,M,d)$
code. We define the capacity of error-correcting codes of minimum
distance $d$ as
\[\ccap(d)=\lim_{n\rightarrow\infty} \frac{\ln A(n,d)}{\ln n!}.\]
It was shown in \cite{BarMaz10} that
\[
\ccap(d)=\begin{cases}
1 & \text{if $d=O(n)$,}\\
1-\epsilon & \text{if $d=\Theta(n^{1+\epsilon})$ with $0<\epsilon<1$,} \\
0 & \text{if $d=\Theta(n^2)$.}
\end{cases}
\]

Turning to systematic codes, let $k(n,d)$ denote the maximum number of
information symbols in systematic codes of length $n$ and minimum
distance $d$. Such codes are $[n,k(n,d),d]$ systematic codes, and have
$k(n,d)!$ codewords. The \emph{capacity} of systematic codes of
minimum distance $d$ is defined as
\[\ccaps(d)=\lim_{n\rightarrow\infty}\frac{\ln k(n,d)!}{\ln n!}.\]
The following theorem shows that systematic codes have the same
capacity as general codes.

\begin{theorem}
The capacity of systematic codes of minimum distance $d$ is
\[\ccaps(d)=\begin{cases}
1 & \text{if $d=O(n)$,}\\
1-\epsilon & \text{if $d=\Theta(n^{1+\epsilon})$ with $0<\epsilon<1$,} \\
0 & \text{if $d=\Theta(n^2)$.}
\end{cases}\]
\end{theorem}
\begin{IEEEproof}
Since systematic codes are a special case of general error-correcting
codes, we naturally have
\[\ccaps(d) \leq \ccap(d).\]
Thus, to prove the claim, all that remains is to prove the other
direction of the inequality.

According to Theorem \ref{theorem2}, there exists an $[n, k,d]$
systematic code if $k$ is the maximum integer that satisfies
\begin{equation}
\label{eq:maxk}
d\binom{k+d}{d}\binom{d+n-k}{n-k}2^n<\frac{n!}{k!}.
\end{equation}
That is because
\begin{multline*}
\sum_{i=1}^{d-1}\binom{k+i-2}{i}\binom{d-i-1+n-k}{n-k}2^{\min{(d-i-1,n-k)}} \leq\\
\leq d\binom{k+d}{d}\binom{d+n-k}{n-k}2^n,
\end{multline*}
for all $n>k\geq 2$ and $d\geq 2$.

For such $k$, we have $k(n,d)\geq k$. For convenience, let us assume
\[\alpha=\lim_{n\rightarrow\infty}\frac{k}{n},\]
is a constant. We also recall the well-known Stirling's approximation,
\[\ln(m!)=m\ln m - O(m).\]
Thus, if $\alpha>0$,
\begin{align*}
\ccaps(d)&=\lim_{n\rightarrow \infty}\frac{\ln k(n,d)!}{\ln n!}\geq \lim_{n\rightarrow \infty}\frac{\ln k!}{\ln n!}\\
&=\lim_{n\rightarrow \infty}\frac{\alpha n \ln (\alpha n)-O(n)}{n\ln n-O(n)}=\alpha.
\end{align*}
To prove the final conclusion, we will show that 
\begin{equation}
\label{eq:toprovealpha}
\alpha\geq \begin{cases}
1 & \text{if $d=O(n)$,}\\
1-\epsilon & \text{if $d=\Theta(n^{1+\epsilon})$,}\\
0 & \text{if $d=\Theta(n^2)$.}
\end{cases}
\end{equation}
We note that the last case is trivial, and so we only have to prove
the first two.

By our choice of $k$ and \eqref{eq:maxk}, we have
\[
d\binom{k+d+1}{d}\binom{d+n-k-1}{n-k-1}2^n\geq\frac{n!}{(k+1)!}.
\]
It follows that
\begin{equation}
\label{equ3_2}
\lim_{n\rightarrow \infty}\frac{\ln \parenv{d\binom{k+d+1}{d}\binom{d+n-k-1}{n-k-1}2^n}}{\ln \parenv{\frac{n!}{(k+1)!}}}\geq 1.
\end{equation}
	
To prove the first case of \eqref{eq:toprovealpha} assume $d=O(n)$.
Again, by Stirling's approximation, \eqref{equ3_2} becomes,
\begin{align*}
1 & \leq \lim_{n\rightarrow \infty}\frac{\ln \parenv{d\binom{k+d+1}{d}\binom{d+n-k-1}{n-k-1}2^n}}{\ln \parenv{\frac{n!}{(k+1)!}}}\\
& = \lim_{n\rightarrow\infty}\frac{O(n)}{n\ln n-\alpha n \ln (\alpha n)-O(n)}.
\end{align*}
Since $\alpha$ is a constant, we must therefore have $\alpha=1$.

For the second case, assume $d=\Theta(n^{1+\epsilon})$ for
$0<\epsilon<1$.  By applying Stirling's approximation to
\eqref{equ3_2}, after some tedious rearranging, we get
\begin{align*}
1 & \leq \lim_{n\rightarrow \infty}\frac{\ln \parenv{d\binom{k+d+1}{d}\binom{d+n-k-1}{n-k-1}2^n}}{\ln \parenv{\frac{n!}{(k+1)!}}}\\
& = \lim_{n\rightarrow\infty}\frac{\epsilon n\ln n - O(n)}{(1-\alpha) n \ln n -O(n)}.
\end{align*}
Thus, $\alpha \geq 1-\epsilon$, as we wanted to show.
\end{IEEEproof}

\section{Systematic Codes in the $\ell_\infty$-Metric}
\label{sec:linf}

We recall that the definition of systematic codes in the
$\ell_\infty$-metric differs from that in Kendall's
$\tau$-metric. Intuitively, in an $[n,k,d]$ systematic code in the
$\ell_\infty$-metric, when taking the first $k$ \emph{coordinates} of
the $k!$ codewords and relabeling the surviving $k$ elements to $[k]$,
we obtain every permutation of $S_k$ exactly once.

The exact capacity for codes in the $\ell_\infty$-metric is
unknown. There is a large gap between the lower and upper bounds on
the size of optimal codes, mainly due to the lack of an asymptotic
expression for the size of balls in this metric. Thus, to evaluate the
parameters of our construction we will compare the rate of the
constructed systematic codes with that of known general codes. Given
an $(n,M,d)$ code $C$ in the $\ell_\infty$-metric, its \emph{rate} is
defined as (see \cite{TamSch10})
\[R(C) = \frac{\log_2 M}{n}.\]
Note that this definition is somewhat different than that for
Kendall's $\tau$-metric (see \cite{BarMaz10}).

We present two constructions for systematic codes, where the first is
adequate for distances $d=O(1)$, and where the second is intended for
the $d=\Theta(n)$ case.

\begin{construction}
\label{con:linfsmalld}
Let $1\leq d\leq n$ be positive integers, and let $1\leq k\leq
\ceilenv{n/d}$ be an integer as well. We denote
\[A_{k,d} = \mathset{1,1+d,1+2d,\dots,1+(k-1)d}.\]
We construct the code
\begin{align*}
C = \{ (f_1,\dots,f_n)\in S_n ~|~ &\mathset{f_1,\dots,f_k}=A_{k,d} ,\\
&f_{k+1}<f_{k+2}<\dots<f_n\}
\end{align*}
\end{construction}

\begin{theorem}
The code $C$ from Construction \ref{con:linfsmalld} is an $[n,k,d]$
systematic code in the $\ell_\infty$-metric.
\end{theorem}
\begin{IEEEproof}
It is immediately evident that $\abs{C}=k!$. Furthermore, every two
distinct codewords in $C$ disagree on at least on of the first $k$
coordinates. Since all entries in the first $k$ coordinates leave a
residue of $1$ modulo $d$, the $\ell_\infty$-distance between distinct
codewords is at least $d$. Finally, the projection onto the first $k$
coordinates is easily seen to provide all possible permutations from
$S_k$ exactly once.
\end{IEEEproof}

The optimal choice of $k$ in Construction \ref{con:linfsmalld} is
obviously $k=\ceilenv{n/d}$, and it provides a code of size
$\ceilenv{n/d}!$. This can be compared with Construction 1 of
\cite{TamSch10} which gives a code of size
$\parenv{\ceilenv{n/d}!}^{n\bmod d}\parenv{\floorenv{n/d}!}^{d-(n\bmod
  d)}$. If we denote the rate of the code from Construction
\ref{con:linfsmalld} by $R$, and the rate of the code from
Construction 1 of \cite{TamSch10} by $R'$, then
\[\frac{R}{R'} = \frac{\log_2(\ceilenv{n/d}!)}
{\log_2\parenv{\parenv{\ceilenv{n/d}!}^{n\bmod d}\parenv{\floorenv{n/d}!}^{d-(n\bmod
  d)}}} \geq \frac{1}{d}.\]

We now turn to provide a construction suited for $d=\Theta(n)$.

\begin{construction}
\label{con:linfbigd}
Let $1\leq d\leq n$ be positive integers. We recall Construction 1 from
\cite{TamSch10}, of an $(n,M,d)$ code,
\[ C'=\mathset{ f\in S_n ~|~ \text{$f(i)\equiv i\pmod{d}$ for all $i$ }},\]
where
\[ M = \abs{C'}=\parenv{\ceilenv{n/d}!}^{n\bmod d}\parenv{\floorenv{n/d}!}^{d-(n\bmod
  d)}.\]
Let $k$ be the largest integer such that $k!\leq \abs{C'}$, and let
$C''$ be the set of all permutations over the set
$\mathset{n+1,n+2,\dots,n+k}$. Assume
\begin{align*}
C' & = \mathset{f'_1,f'_2,\dots,f'_{\abs{C'}}} \\
C'' & = \mathset{f''_1,f''_2,\dots,f''_{k!}}.
\end{align*}
We now construct the code
\[C=\mathset{ f''_i\|f'_i ~|~ 1\leq i\leq k!},\]
where $\|$ denotes vector concatenation.
\end{construction}
\begin{theorem}
The code $C$ from Construction \ref{con:linfbigd} is an $[n+k,k,d]$ systematic
code in the $\ell_\infty$-metric.
\end{theorem}
\begin{IEEEproof}
The code is obviously of size $k!$, and by construction, the
projection onto the first $k$ coordinates gives all possible
permutations exactly once. Since $C'$ is a code with minimum distance
$d$ (see \cite{TamSch10} for proof), the code $C$ also has minimum
distance of $d$ in the $\ell_\infty$-metric.
\end{IEEEproof}

We now turn to analyze the asymptotic rate of the code from
Construction \ref{con:linfbigd}. Assume $d=\delta n$, where $\delta$
is a constant, $0<\delta <1$. By our choice of $k$, we have
\begin{equation}
\label{eq:ratio}
\frac{1}{k+1}\abs{C'}\leq \abs{C}=\abs{C''} =k! \leq \abs{C'}.
\end{equation}
Let $R$ denote the rate of $C$, and $R'$ denote the rate of $C'$, i.e.,
\begin{align*}
R & = \frac{\log_2 k!}{n+k},\\
R' & = \frac{\log_2\abs{C'}}{n}.
\end{align*}
Thus, by \eqref{eq:ratio},
\[\parenv{1-\frac{\log_2 (k+1)}{\log_2\abs{C'}}} \frac{n}{n+k} \leq \frac{R}{R'}\leq \frac{n}{n+k}.\]
Since $1\leq k\leq n$, while (see \cite{TamSch10})
\[\abs{C'}\geq 2^{(1-\delta)n}\]
we have
\[\lim_{n\to\infty} \frac{R}{R'}=\lim_{n\to\infty}\frac{n}{n+k}.\]

At this point we need to bound $k$, and we contend that 
\[k\leq \frac{n}{\log_2\log_2 n}.\]
Let us assume, for $k=\ceilenv{n / \log_2\log_2 n}$,
that we have
\[k! \leq \abs{C'}.\]
We easily see that
\[\abs{C'} \leq \parenv{\ceilenv{\frac{n}{d}}!}^d=
\parenv{\ceilenv{\frac{1}{\delta}}!}^{\delta\ceilenv{1/\delta}n}=\alpha^n\]
for some constant $\alpha>1$.

On the other hand, we recall the well known bound
\[m! \geq \parenv{\frac{m}{e}}^m,\]
which holds for all positive integers $m$. Thus,
\[k!=\ceilenv{n/\log_2 \log_2 n}! \geq \parenv{\frac{n}{e\log_2\log_2 n}}^{\frac{n}{\log_2\log_2 n}}.\]
If indeed $k!\leq \abs{C'}$ then necessarily
\[ \parenv{\frac{n}{e\log_2\log_2 n}}^{\frac{n}{\log_2\log_2 n}} \leq \alpha^n,\]
and taking $\log_2$ of both sides gives us
\[ \frac{n}{\log_2\log_2 n}\log_2 \parenv{\frac{n}{e\log_2\log_2 n}}
\leq n\log_2 \alpha.\]
However, this last inequality certainly does not hold for large enough $n$.
It therefore follows that indeed
\[k\leq \frac{n}{\log_2\log_2 n}.\]
Finally,
\[\lim_{n\to\infty} \frac{R}{R'} =\lim_{n\to\infty}\frac{n}{n+k}
\geq \lim_{n\to\infty}\frac{n}{n+\frac{n}{\log_2\log_2 n}}
= 1.\]

Essentially, when $d=\Theta(n)$, we constructed systematic codes with
the same asymptotic rate and minimum distance as the non-systematic
codes appearing in \cite{TamSch10}, which are currently the best codes
known asymptotically.  Furthermore, the construction we presented can
work with any other non-systematic error-correcting code, provided it
has an exponential size when $d=\Theta(n)$.

\section{Conclusion}
\label{sec:conclusion}

In this paper, we studied systematic error-correcting codes for rank
modulation under two metrics: Kendall's $\tau$-metric, and the
$\ell_\infty$-metric. In the former, we presented several
constructions, explicit and algorithmic, and found the capacity of
systematic codes. Efficient encoding and decoding schemes were also
discussed. In the latter, two constructions were given, one of them
asymptotically attaining the same rate as the best construction
currently known in this metric.

Some open questions remain. In Kendall's $\tau$-metric we still do not
know, given $n$ and $d$, what is the largest $[n,k,d]$ systematic code
possible. We are also interested in the question of whether systematic
perfect codes (or even general perfect codes) exist. In the
$\ell_\infty$-metric, we are still missing tight bounds, even
asymptotically, on the parameters of general codes, as well as for
systematic codes.

\bibliographystyle{IEEEtranS}
\bibliography{allbib}

\end{document}